\documentclass[aip,graphicx,reprint]{revtex4-1}

\usepackage{color}
\usepackage{graphicx}
\usepackage{siunitx}
\usepackage{lineno}
\usepackage{hyperref}

\begin{document}

\preprint{Ver.~proof}
…

\title{Fast X-ray detector system with simultaneous measurement of timing and energy for a single photon} 



\author{T.~Masuda}
 \email[]{masuda@okayama-u.ac.jp}
\affiliation{Research Institute for Interdisciplinary Science, Okayama University, Okayama 700-8530, Japan}

\author{S.~Okubo}
\affiliation{Graduate School of Natural Science and Technology, Okayama University, Okayama 700-8530, Japan}

\author{H.~Hara}
\affiliation{Research Institute for Interdisciplinary Science, Okayama University, Okayama 700-8530, Japan}

\author{T.~Hiraki}
\affiliation{Research Institute for Interdisciplinary Science, Okayama University, Okayama 700-8530, Japan}

\author{S.~Kitao}
\affiliation{Research Reactor Institute, Kyoto University, Osaka 590-0494, Japan}

\author{Y.~Miyamoto}
\affiliation{Research Institute for Interdisciplinary Science, Okayama University, Okayama 700-8530, Japan}

\author{K.~Okai}
\affiliation{Graduate School of Natural Science and Technology, Okayama University, Okayama 700-8530, Japan}

\author{R.~Ozaki}
\affiliation{Graduate School of Natural Science and Technology, Okayama University, Okayama 700-8530, Japan}

\author{N.~Sasao}
\affiliation{Research Institute for Interdisciplinary Science, Okayama University, Okayama 700-8530, Japan}

\author{M.~Seto}
\affiliation{Research Reactor Institute, Kyoto University, Osaka 590-0494, Japan}

\author{S.~Uetake}
\affiliation{Research Institute for Interdisciplinary Science, Okayama University, Okayama 700-8530, Japan}

\author{A.~Yamaguchi}
\affiliation{Quantum Metrology Laboratory, RIKEN, Saitama 351-0198, Japan}

\author{Y.~Yoda}
\affiliation{Japan Synchrotron Radiation Research Institute, Hyogo 679-5198, Japan}

\author{A.~Yoshimi}
\affiliation{Research Institute for Interdisciplinary Science, Okayama University, Okayama 700-8530, Japan}

\author{K.~Yoshimura}
\affiliation{Research Institute for Interdisciplinary Science, Okayama University, Okayama 700-8530, Japan}

\date{\today}

\begin{abstract}
  We developed a fast X-ray detector system for nuclear resonant scattering (NRS) experiments. Our system employs silicon avalanche photo-diode (Si-APD) as a fast X-ray sensor. The system is able to acquire both timing and energy of a single X-ray photon simultaneously in a high rate condition, $10^6$ counts per second for one Si-APD. The performance of the system was investigated in SPring-8, a synchrotron radiation facility in Japan. Good time resolution of \SI{120}{ps} (FWHM) was achieved with a slight tail distribution in the time spectrum by a level of $10^{-9}$ at \SI{1}{ns} apart from the peak. Using this system, we successfully observed the NRS from the 26.27-keV level of mercury-201, which has a half-life of \SI{630\pm50}{ps}. We also demonstrated the reduction of background events caused by radioactive decays in a radioactive sample by discriminating photon energy.
\end{abstract}


\maketitle 

\section{Introduction}
\label{sec:intro}
Recent high intensity X-ray beams obtained from synchrotron radiation (SR) in electron storage rings enable us to study nuclear and material properties by using nuclear resonant scattering (NRS).\cite{rohlsberger,Sturhahn2004} \footnote{The term NRS has several definitions among literature. In this study, we use NRS as a broad term that includes both coherent and incoherent, and also both elastic and inelastic.} Combined with a method of M\"ossbauer spectroscopy, SR-based NRS studies can analyze not only static but also dynamical properties of materials.\cite{Seto1995} Such a method has attracted great interest from a wide range of research areas including physics, material and life science, and earth science.\cite{[{See, for example: }]Lubbers2000,*Mao914,*Keppler1997,*Wong2013} Actually, dedicated beamlines for NRS research are equipped in present third-generation SR facilities.\cite{Yoda2001715,Alp1994,Ruffer1996} The usefulness of the method can be extended further by higher brightness of X-ray beams achievable at X-ray free electron laser facilities.\cite{xfel}

 The SR-based NRS technique may be applicable to a wide variety of nuclides, since the energy of X-ray beams is tunable up to $\sim$\SI{100}{keV}. However, so far nuclides in which lifetimes of nuclear states are relatively long ($\gtrsim$ \SI{1}{ns}) have been studied. In contrast, very few nuclides with shorter lifetime have been studied to date mainly due to technical difficulties in measuring short lifetime of resonances. The detector system would require good time resolution to distinguish small NRS signals with a short lifetime ($\sim$ns) from a huge amount of prompt scattering of the incident X-rays, e.g., Rayleigh, Compton scattering, and X-ray fluorescence. The small signal--to--prompt-scattering ratio requires the detector to be tolerant against a high counting rate in order to obtain enough signal counts without saturation by high-rate prompt scattering. The good time resolution is also desired for coherent forward scattering in order to resolve fine interference patterns. There are reports on fast detectors showing applications in measurement of the fine interference pattern of dysprosium-161 samples.\cite{Baron2001,Baron2002}

Another point to be mentioned is a difficulty in the NRS measurement of radioactive nuclides. Constant radiation from radioactive decays would reduce the signal-to-noise ratio. NRS of potassium-40, which is a radioactive nuclide, has been observed\cite{PhysRevLett.84.566} by virtue of its very low specific radioactivity of \SI{2.58E5}{Bq/g}.\cite{toi} However, most of nuclides having higher radioactivity, e.g., thorium-229 ($^{229}$Th, \SI{7.85E9}{Bq/g}\cite{toi}) of which the NRS has been proposed,\cite{Yoshimura2014,PhysRevC.61.064308} have not been studied to date due to a poor signal-to-noise ratio. If one can obtain the energy information of an X-ray photon, one can filter out background radiation events from acquired data. That is because a distribution of the deposit energy due to radioactivities is generally different from that due to the NRS. This is one of the essential techniques to search for such small signals. A crucial point in this technique is to combine energy measurement with high rate capability.

As a fast single X-ray photon detector, a silicon avalanche photo-diode (Si-APD)\cite{Kishimoto1991} is widely used. Although there are many fast X-ray detectors,\cite{Rochau2006,Lihong2008,Inderbitzin2012} Si-APDs are suitable for the NRS measurement due to its large sensitive area and low dark count rate. In particular, a short tail in its time response to the prompt scattering is especially important because of the small signal--to--prompt-scattering ratio, e.g., $10^{-7}$--$10^{-8}$ in this work. Several reports describe the importance of short tail components and demonstrate short tails.\cite{PhysRevB.72.140301,Kishimoto1994}
Detailed specifications have been systematically studied.\cite{Baron2006}

We have developed a fast X-ray detector system. This system mainly focuses on fast time response, not only good overall time resolution but also a short tail. 
 It is also able to acquire the X-ray photon energy in a high-rate condition by developing a dedicated fast circuit; this feature enables us to differentiate NRS signals from background by using the photon energy information in addition to the time information. 
  This system is basically designed with an aim to measure incoherent scattering from NRS with very short lifetimes and small excitation cross sections, which are quite difficult to be measured. The realized performance can be utilized for other applications desiring fast time response or energy information.
  The system consists of commercially available Si-APDs, fast frontend circuits, and time-to-digital converters (TDCs).
 We investigated its performance in SPring-8, a synchrotron radiation facility in Japan. We successfully observed NRS signals from the 26.27-keV level of mercury-201 ($^{201}$Hg), the lifetime of which is shorter than \SI{1}{ns}. We also demonstrated the applicability of energy information in measurement of radioactive nuclides, optimizing experimental sensitivity after the data acquisition. In this paper, each component constituting the detector system is described in Sec.~\ref{sec:detector}. The experimental results in SPring-8 are reported in Sec.~\ref{sec:result}. Section~\ref{sec:conclusion} concludes the paper.

\section{Detector system}
\label{sec:detector}

\begin{figure}
 \includegraphics[width=8.5cm]{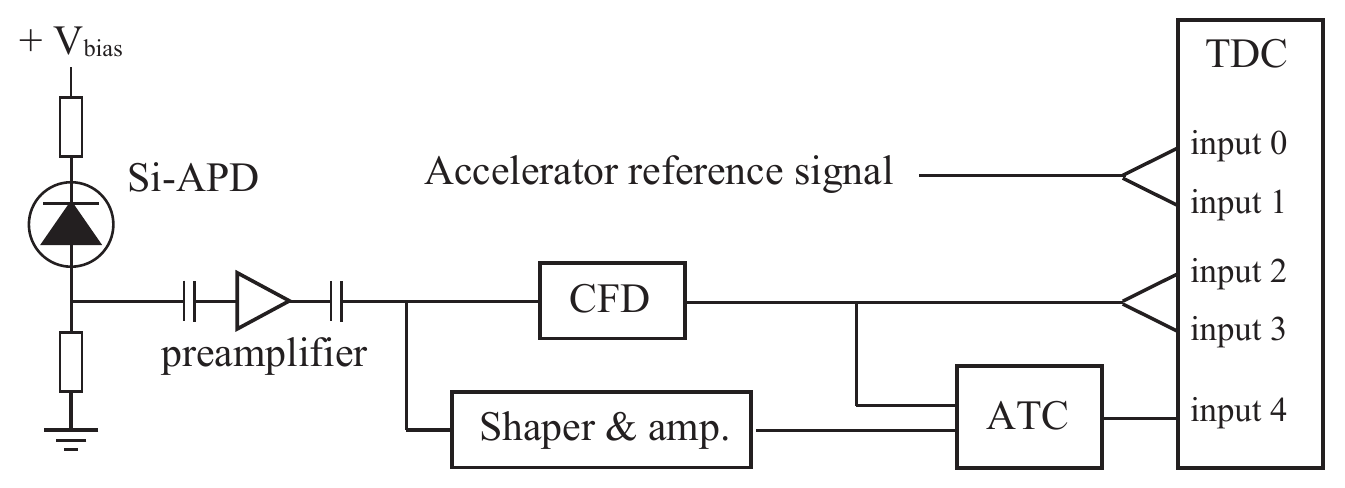}
 \caption{\label{fig:diagram}Block diagram of a channel in the system. Si-APD: silicon avalanche photo-diode; CFD: constant fraction discriminator; ATC: amplitude-to-time convertor; TDC: time-to-digital convertor.}
\end{figure}

 The system consists of six sets of a Si-APD, frontend fast pulse-processing circuits, and a TDC. The block diagram of one channel is shown in Fig.~\ref{fig:diagram}. The accelerator reference signal provides a start signal and the Si-APD output signal is used as a stop signal after being converted to a logic pulse by a constant-fraction discriminator (CFD). An amplitude-to-time convertor (ATC) output is used to measure a pulse height; the time difference between the CFD output and the ATC output is proportional to the pulse height. All the accelerator reference signal, CFD output signal, and ATC output signal are fast Nuclear Instrumentation Module (fast-NIM) logic signals.
 
\begin{figure}
\includegraphics[width=6cm]{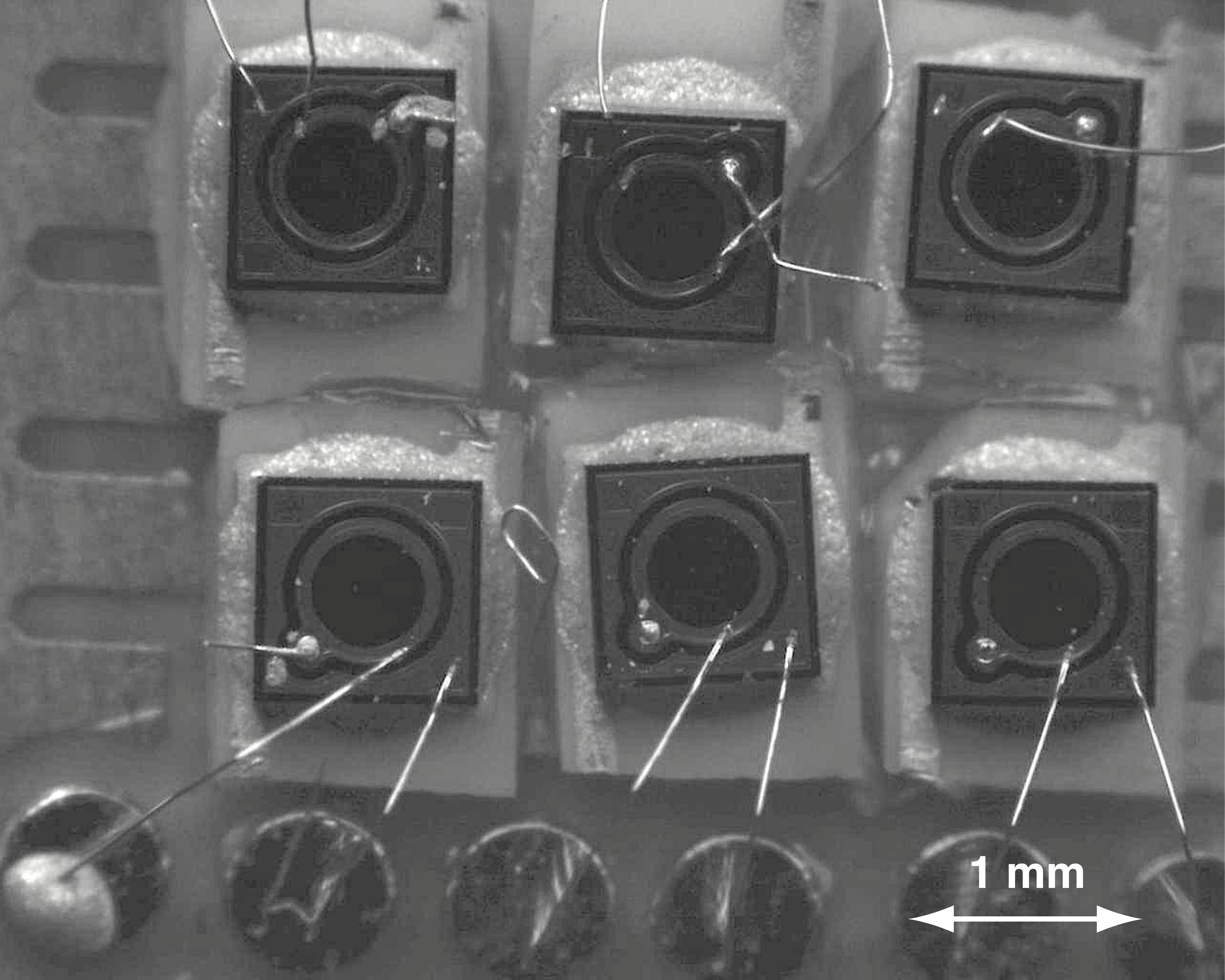}
\caption{\label{fig:dice}Photograph of the Si-APD $2 \times 3$ array. Diameter of the sensitive area is \SI{0.5}{mm}. Aluminum wires connect each electric pad of the Si-APD to a metal pad on the substrate.}
\end{figure}

\begin{figure}
\includegraphics[width=7.5cm]{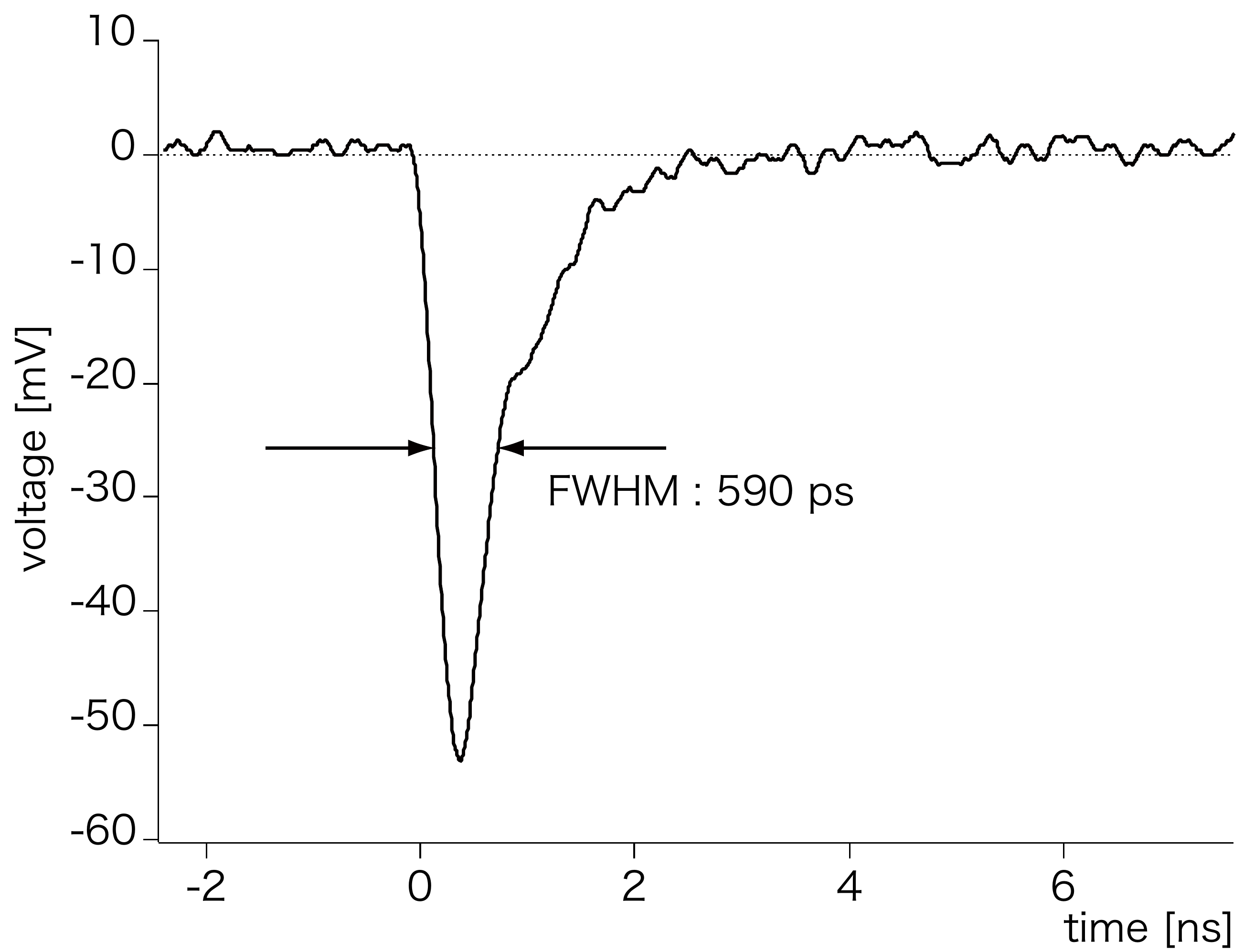}
\caption{\label{fig:pulse}Pulse shape of the preamplifier output obtained with a cadmium-109 X-ray radioactive source. It is digitized in 40-gigahertz sampling rate with a 3.5-GHz bandwidth oscilloscope (Tektronix, DPO7354C). Reverse bias voltage of \SI{150}{V} is applied to the Si-APD.}
\end{figure}

 The Si-APD (Hamamatsu Photonics, S12053-05) has a sensitive area of \SI{0.5}{mm} in diameter. It is adopted because of the fast time response due to its thin depletion layer of \SI{10}{\micro m} which is estimated from the datasheet. The thin depletion layer causes the absorption efficiency to be low: 2\% for 15-keV photons and 0.2\% for 30-keV photons. We built a six-channel Si-APD array, shown in Fig.~\ref{fig:dice}, to compensate the detection efficiency by increasing the geometrical acceptance. We dismounted six APD silicon chips from commercial metal-can packages and then mounted them on a lab-build substrate.
 Each Si-APD chip is directly connected to a preamplifier (Mini Circuits, RAM-8A+) whose gain and 3-dB cutoff frequency were measured to be \SI{+32}{dB} and \SI{453}{MHz}, respectively. A typical output signal pulse of the preamplifier for an X-ray photon is shown in Fig.~\ref{fig:pulse}. The pulse width (FWHM) and the rise time from 10\% to 90\% are \SI{0.59}{ns} and \SI{0.27}{ns}, respectively. 
 
\begin{figure}
\includegraphics[width=8cm]{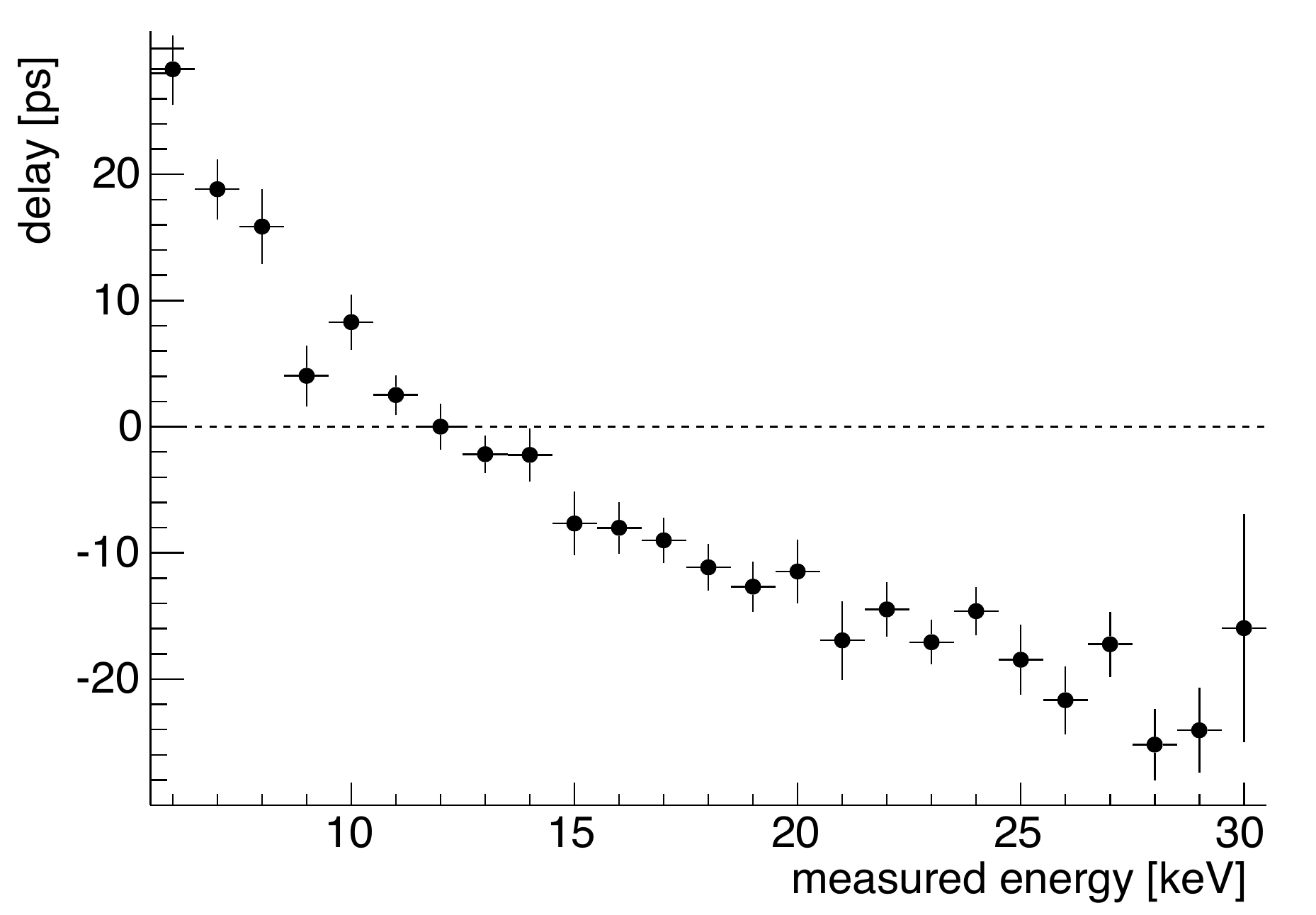}
\caption{\label{fig:cfd}Pulse height dependence of the CFD output. The horizontal axis is calibrated from the pulse height to the measured energy. The origin of the vertical axis is set at the \SI{12}{keV} point. It is obtained with a cadmium-109 X-ray radioactive source.}
\end{figure}

\begin{figure}
\includegraphics[width=8.5cm]{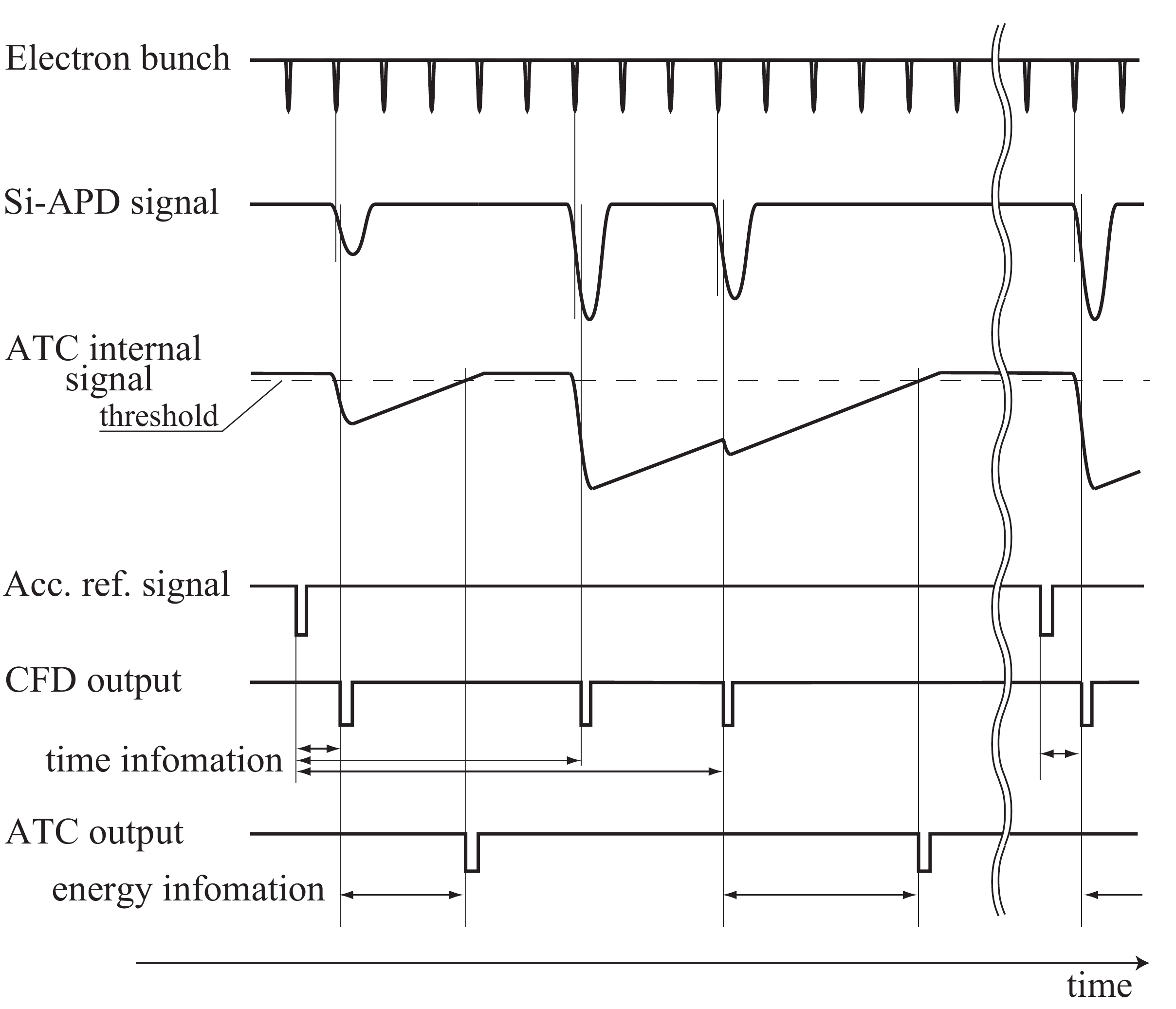}
\caption{\label{fig:stream}Data stream for one channel. The most upper line shows timing of electron-bunchs circulating in a storage ring. The second line shows analog pulses sent from a Si-APD and the third line shows internal analog signals from the peak-hold and discharge circuits in the ATC. The other lines show logic pulses which are recorded by the TDC. The horizontal axis is not scaled.}
\end{figure}

 The preamplifier output signal is sent to the CFD and a pulse shaping circuit. The CFD is used to suppress time-walk effect\cite{leo} due to the rise time of the analog pulses. In addition, putting the CFD just after the preamplifier can eliminate the pulse shape relaxation during transmission. The CFD consists of ultrafast comparators (Analog devices, ADCMP581). Figure~\ref{fig:cfd} shows pulse height, corresponding to measured energy, dependence of the timing delay in the CFD. The deviation is about $\pm 25$\,ps which is small enough for the timing measurement. The CFD also provides a gate signal to the following ATC. The pulse shaping circuit expands a pulse width and sends the pulse to the ATC. To maintain the high rate capability of the system, we developed a fast ATC. The ATC consists of a peak-hold circuit and a constant discharge circuit. Figure~\ref{fig:stream} shows a time diagram of the ATC operation together with the data stream of a channel. It outputs a logic pulse whose time interval from the input signal is proportional to the pulse height. The conversion time is less than \SI{10}{ns/keV}. If the next signal comes in during the conversion, the output from the peak-hold circuit is updated and the previous signal is ignored.

 A multi-stop TDC (FAST ComTec, MCS6), which is a multiple-event time digitizer that equips six input channels with 100-ps per time bin, is used for recording both time and energy information of each X-ray photon. We split both accelerator reference signal and CFD output pulse into two input channels each with 50-ps difference so as to achieve 50-ps per time bin effectively. The TDC digitizes an arrival time of one pulse to 54-bit data and sends it to a computer via an Universal Serial Bus (USB) 2.0 interface. The maximum readout speed of the TDC is $\sim$\SI{35}{MB/s}. The rate capability of our system is approximately $1 \times 10^6$ counts per second (cps) for one Si-APD; it is limited by the TDC readout speed.
 
 The recorded data are converted to timing and energy information for each signal using software. The data stream in time domain is shown in Fig.~\ref{fig:stream}. 
 The time information is converted from the time difference between a CFD output pulse and the last accelerator reference signal which synchronizes the accelerator RF cycle. Because one reference signal comes in once every a certain number (203 in this work) of electron bunches, which depends on accelerator operation modes, the time offsets between an accelerator reference signal and timing of an electron bunch should be subtracted for each bunch. It is done after the data acquisition.
 The energy information is converted from the time difference between an ATC output pulse and the last CFD output pulse. If the count rate is too high, the number of ATC pulses is less than that of CFD pulses because the conversion time of the ATC is longer. The CFD pulses that do not have a corresponding ATC pulse are ignored. The above processes are done for each TDC individually and finally all converted data are gathered in one data server.
 
\section{System integration in SPring-8}
\label{sec:result}
 This system has various advantages over previous systems: energy measurement, fast time response, small tail, simultaneous energy--time measurement, and fast scalable data acquisition. The \textit{in situ} performance of the integrated system was investigated at the BL09XU beamline of SPring-8.

\subsection{Experimental setup}
\begin{figure}
\includegraphics[width=8cm]{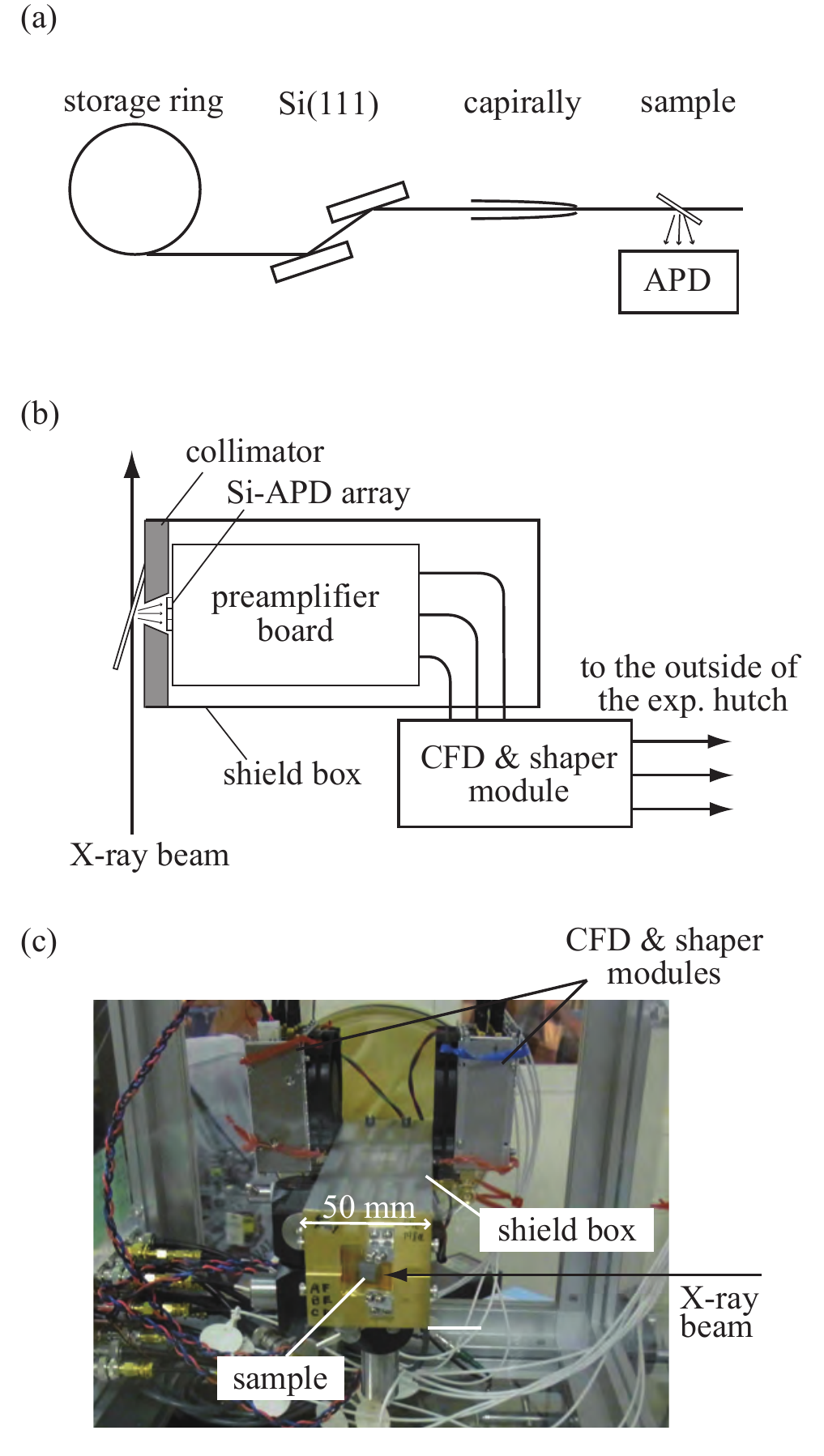}
\caption{\label{fig:setup}(a) Schematic view of the experimental setup at SPring-8. (b) Close up view at the sample and the Si-APD array detector. One CFD and shaper module is drawn and the other one is omitted. (c, color online) Photo of the detector frontend.}
\end{figure}

 The experimental set up is shown in Fig.~\ref{fig:setup}. The operation mode of SPring-8 was a 203 electron-bunch mode in which the 203 bunches are equally spaced with a time interval of \SI{23.6}{ns} corresponding to \SI{42.4}{MHz} radiation. The ring current was \SI{100}{mA}. A temporal width (FWHM) of an electron bunch in the storage ring is $\sim$\SI{35}{ps}. 
 An accelerator reference signal which is provided for every 203 bunches was sent to the TDC  as shown in Fig.~\ref{fig:diagram} so as to synchronize the accelerator and the detector system. A Si(111) double-crystal monochromator, the bandwidth of which is about \SI{3.4}{eV} (FWHM), was used to adjust energy of the X-ray beam. The available photon yield is in the order of $10^{13}$~photons/s at the downstream of the monochromator and is adjusted by using a X-Y slit in the beamline. The X-ray beam was focused to a spot size of $0.13 \times 0.11$\,\si{mm^2} by a tapered glass capillary (HORIBA, 2014SP13). The spot size is defined as the standard deviation of the cross section.
 
 The Si-APD array and the preamplifiers were housed in an aluminum shield box so as to prevent electric noise. The Si-APDs were operated in the linear mode. Reverse bias voltage of +\SI{150}{V} was applied to all Si-APDs in which the nominal gain is 50, and their gain deviation was found to be only 10\% device-by-device. A brass tapered collimator limited the optical path between samples and the Si-APD sensitive areas so that background due to stray X-ray photons was suppressed. The CFDs and shaper modules were directly mounted on the side of the shield box. The CFD output pulses were transferred to the downstream system located at the outside of the experimental hutch via $>$10-m coaxial cables.
 We used three samples for irradiation: a Cu sample, a 0.1-mm-thick copper (Cu) plate; a Hg sample, $\sim$4-mg natural abundant mercury-sulfide (HgS) powder which is pressed at \SI{2}{MPa} to be a 3-mm-diameter plate; and a Th sample, $\sim$4-\si{\micro g} thorium-229 hydroxide [$^{229}\mathrm{Th(OH)}_4$] deposited as a 1.5-mm-diameter circle on a polypropylene sheet (Eichrom, RF-100-25PP01). The radioactivity of the Th sample was \SI{25}{kBq}. The Hg and Th samples were covered with 0.1-mm-thick beryllium windows.

\subsection{Results of pulse height property}
\begin{figure}
\includegraphics[width=8.5cm]{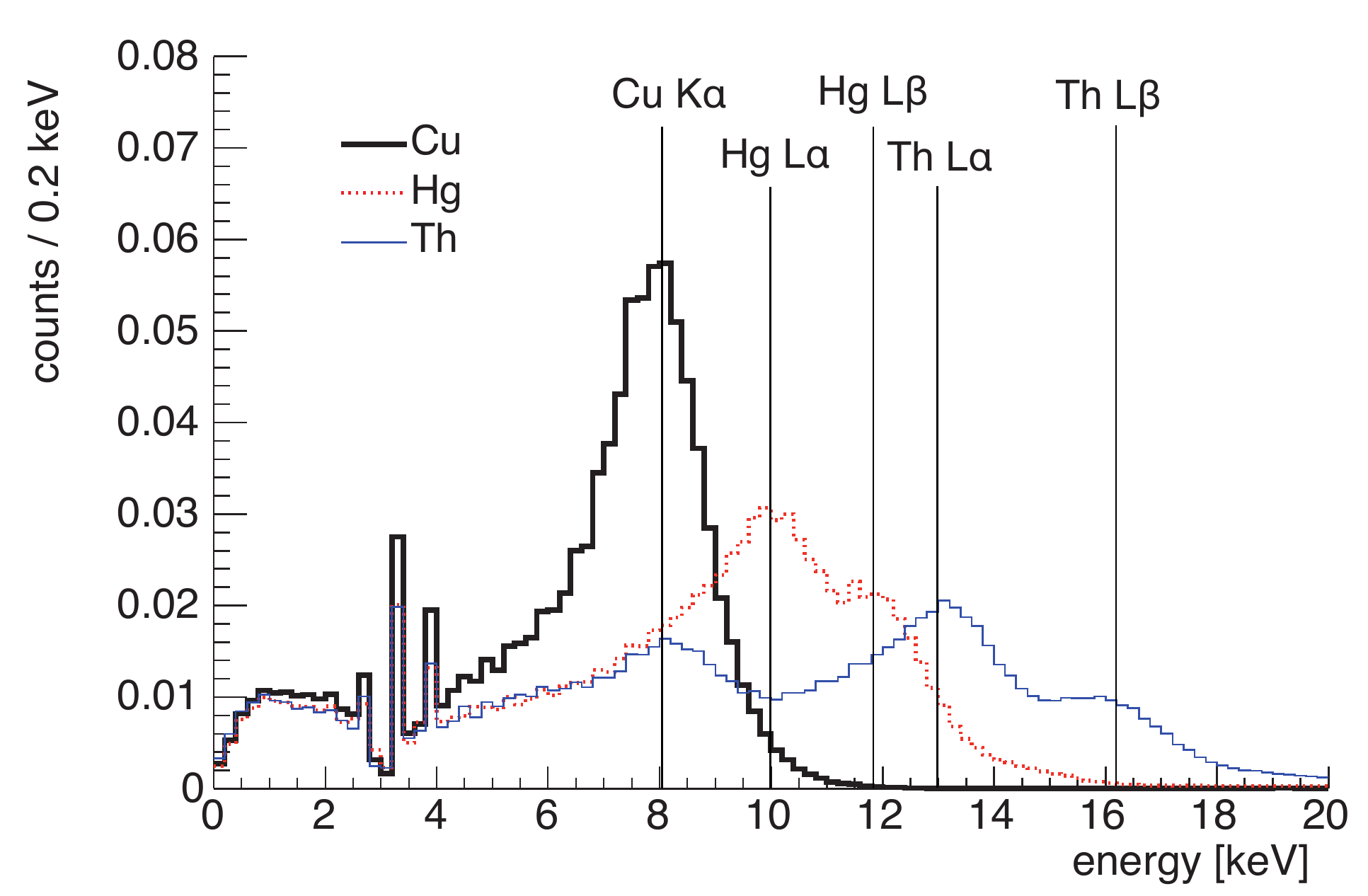}
\caption{\label{fig:pulseheight}(Color online) Energy spectra obtained by using one Si-APD channel. Black bold (red dotted, and blue solid) histogram shows the spectrum of the Cu (Hg, and Th) sample. Each spectrum is normalized so that the integrated number of counts is 1. The five vertical lines indicate the energies of the characteristic peaks in the literatures.\cite{PhysRevA.56.4554,RevModPhys.75.35,PhysRevA.61.012507} The 3--4\,\si{keV} region is disturbed by electric noise (see text).}
\end{figure}

 Irradiating each sample with the X-ray beam, we measured the energy distribution of scattered X-rays. The measured energy spectra obtained with the samples are shown in Fig.~\ref{fig:pulseheight}. They show sufficiently good linearity and energy resolution which enable identification of individual characteristic X-ray peaks: K$\alpha$ (\SI{8.0}{keV}\cite{PhysRevA.56.4554}) of Cu, L$\alpha$ (\SI{10.0}{keV}\cite{RevModPhys.75.35}) and L$\beta$ (\SI{11.8}{keV}\cite{RevModPhys.75.35}) of Hg, and L$\alpha$ (\SI{13.0}{keV}\cite{PhysRevA.61.012507}) and L$\beta$ (\SI{16.2}{keV}\cite{PhysRevA.61.012507}) of Th.
 The width of the Cu K$\alpha$ peak is 23\% (FWHM). We calibrated the pulse height to energy by linear fitting using the five peaks for each channel. We use the calibrated energy instead of the pulse height in the latter part of this paper.
 
 Each spectrum has a tail structure extending to the low energy region. The behaviour is generally explained by contribution of X-ray photons that are absorbed in the multiplication region. The gain in that case depends on the position along the thickness.\cite{Yatsu2006}
 The region at 3--4\,keV in each spectrum is disturbed due to digital electric noise in the ATC circuit boards. It is not a matter because it locates outside the region of interest.
 The peak at \SI{8}{keV} in the spectrum of the Th sample is due to characteristic X-rays of copper or zinc from the brass collimator in front of the Si-APD array.

\subsection{Results of temporal profile}
\begin{figure}
\includegraphics[width=8.5cm]{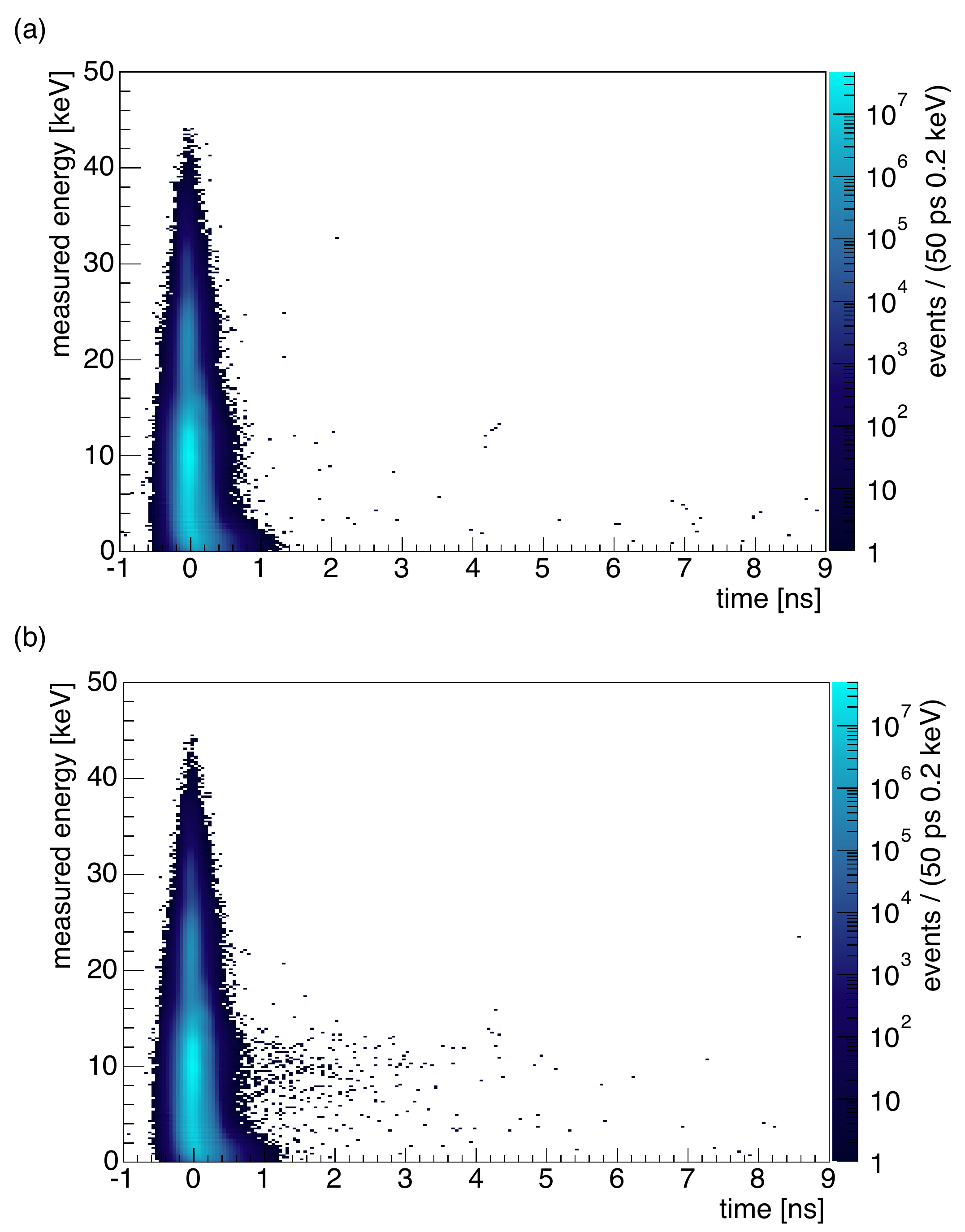}
\caption{\label{fig:hgsscatter}(Color online) Two-dimensional histograms of the measured energy and timing obtained with the Hg sample. The horizontal axis is shifted so that the peak of the prompt scattering locates at the origin. (a) The spectrum at the off-resonance energy. (b) The spectrum at the on-resonance energy. }
\end{figure}

 The NRS from the second-excited 26.27-keV state in $^{201}$Hg was measured to demonstrate the feasibility of the system. Figures~\ref{fig:hgsscatter}(a) and \ref{fig:hgsscatter}(b) show two-dimensional (2D) histograms of the measured energy and timing accumulated at the off-resonance and on-resonance X-ray energy, respectively. The energy difference between the two conditions is \SI{17}{eV}, five times as large as the bandwidth of the monochromator. The photon flux at the target was $2\times 10^{11}$\,photons/s. The counting rate of NRS signals was \SI{0.7}{cps}. The total counting rate of six channels was \SI{4.3E6}{cps} and the maximum counting rate per channel among six was \SI{1.1E6}{cps}. The data of all the six channels integrated in \SI{1000}{s} were filled in both the histograms. 
 The photons of the NRS signals are mainly X-ray fluorescence following the internal conversion, in which the nuclear excitation energy takes a surrounding electron off, because of its large internal conversion factor of 71.6(6).\cite{KONDEV2007365} Therefore, both energy distributions of the NRS signals and the prompt scattering locate around the X-ray characteristic peaks. The existence of well above \SI{40}{keV} energy irrespective of incident energy of \SI{26.27}{keV} is mainly due to pile up of multi X-ray hits.
 
\begin{figure}
\includegraphics[width=8.4cm]{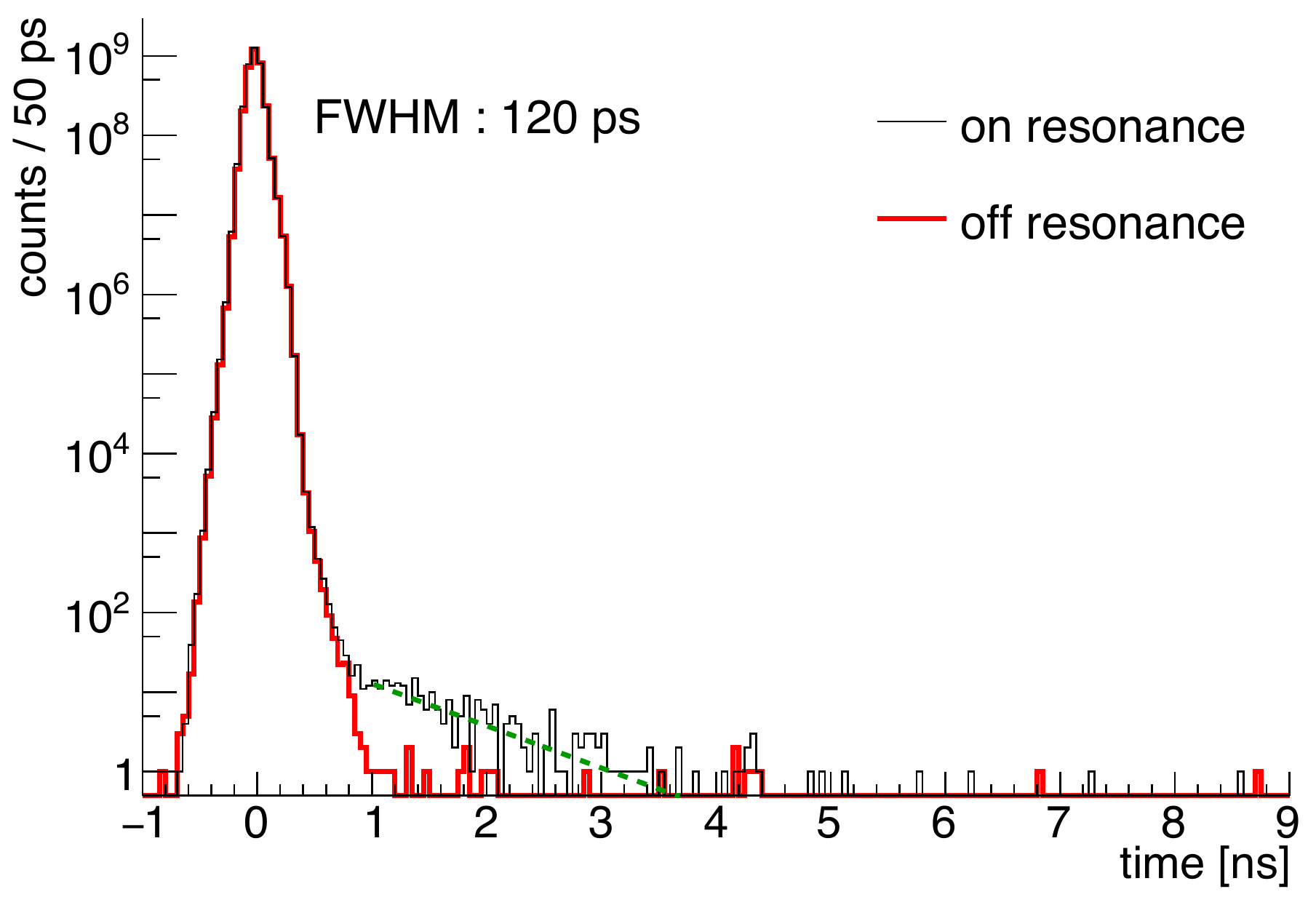}
\caption{\label{fig:hgsspectrum}(Color online) Temporal spectra of the NRS measurement obtained with the Hg sample. The red bold histogram is the spectrum at  the off-resonance energy; the black solid histogram, the on-resonance energy. The green dashed line is a fit result with an exponential function.}
\end{figure}

 Figure~\ref{fig:hgsspectrum} shows the temporal spectra with the energy selection of higher than \SI{5}{keV}.  
 The peak of the prompt scattering represents good time resolution of \SI{120}{ps} (FWHM) including the temporal width of the bunches; furthermore, the tail rapidly decreases down to $10^{-9}$ of maximum within \SI{1}{ns} from the peak. The shortness of the tail enables us to observe the slope of the fast NRS signals clearly. The half-life and the statistical uncertainty are estimated to be \SI{568 \pm 72}{ps} with a reduced chi-square of 1.02 by an exponential fitting from 1 to \SI{4}{ns}. It is consistent with the previous reported value of \SI{630 \pm 50}{ps}.\cite{Schuler1983}

The obtained time resolution of \SI{120}{ps} is one of the best for array type Si-APD detectors. In addition, the system realizes the unprecedented short tail of \SI{1}{ns} at $10^{-9}$ of maximum, compared to the level of \SI{2}{ns} at $10^{-5}$ in the previous $^{201}$Hg NRS measurement.\cite{PhysRevB.72.140301} The short tail is also useful for the temporal profile measurement of X-ray bunches. This application has been reported and the detector performance in the report was \SI{2}{ns} at $10^{-8}$ of maximum.\cite{Kishimoto1994}

\subsection{Simultaneous measurement of energy and timing}
 The simultaneous acquisition of the energy and timing, and post analysis availability are useful for measuring the NRS of radioactive nuclides. Because experimental sensitivity generally depends on both the signal--to--noise ratio and the absolute number of signals, detail studies on the correlation between energy and timing after the data acquision can maximize the sensitivity finely. These functions are really effective in searches for rare signals with uncertain cross section and lifetime such as $^{229}$Th because proper energy threshold and time window cannot be determined before or during measurement. We demonstrated the applicability of the system using the Th sample. The X-ray energy in this study was set to \SI{29.19}{keV} which is the expected resonant energy of the second-excited state in $^{229}$Th.\cite{PhysRevLett.98.142501} 
 
\begin{figure}
\includegraphics[width=8.5cm]{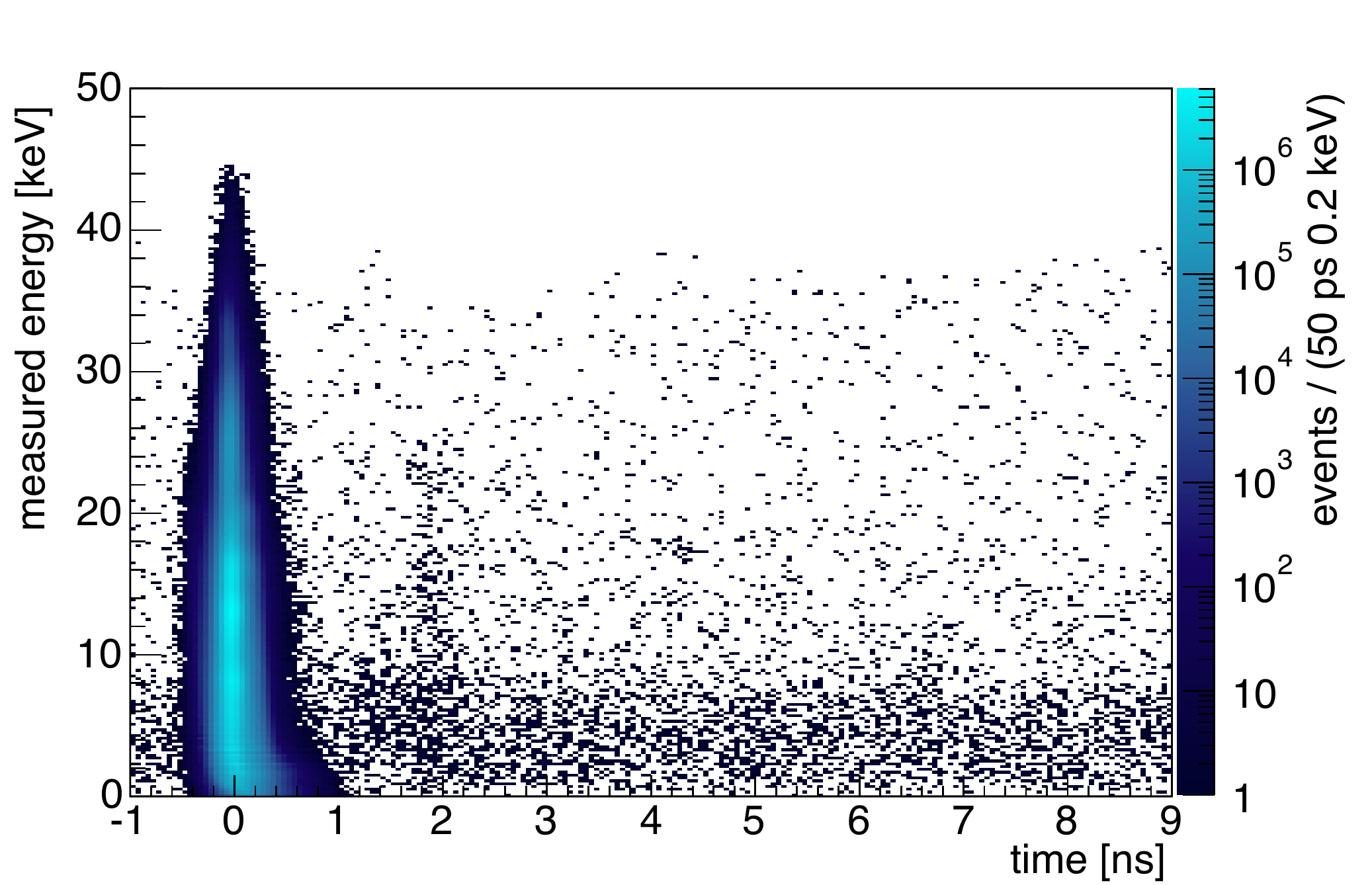}
\caption{\label{fig:thscatter}(Color online) Two-dimensional histogram of the measured energy and timing obtained with the Th sample. The huge peak at the origin of the horizontal axis consists the prompt scattering and the sparsely distributed constant component along the horizontal axis is due to the radioactivity of the Th sample.}
\end{figure}

  A 2D histogram of measured energy and timing is shown in Fig.~\ref{fig:thscatter}. It is accumulated in a \SI{200}{s} run. The counting rate of the run was \SI{4.2E6}{cps} for all channels and the maximum counting rate per channel was \SI{1.4E6}{cps}. There is constant background due to the radioactivity of $^{229}$Th and its daughter elements. The other component located at \SI{2}{ns} might be caused by stray scattering X-rays from surrounding structures, because the volume of the Th sample is quite small in comparison to the Hg sample and the stray X-rays enter the Si-APD sensitive areas in relatively easy paths.
  
\begin{figure}
\includegraphics[width=8.5cm]{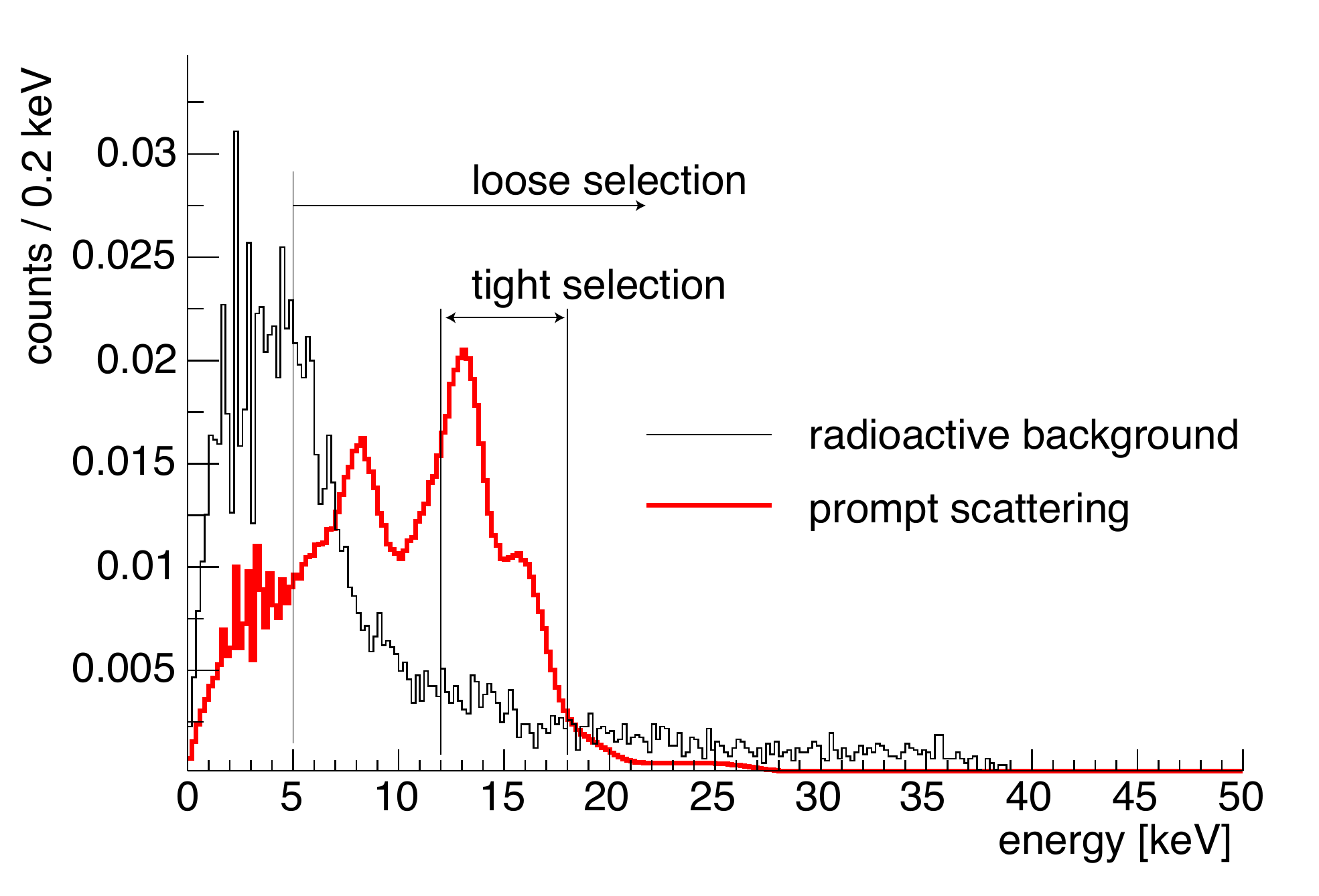}
\caption{\label{fig:thenergy}(Color online) Energy spectra of the radioactive background (black solid) and of the prompt scattering (red bold). Both histograms are normalized so that their integrated areas are 1. The two selections represented as arrows are used in Fig.~\ref{fig:thspectrum}.}
\end{figure}

\begin{figure}
\includegraphics[width=8cm]{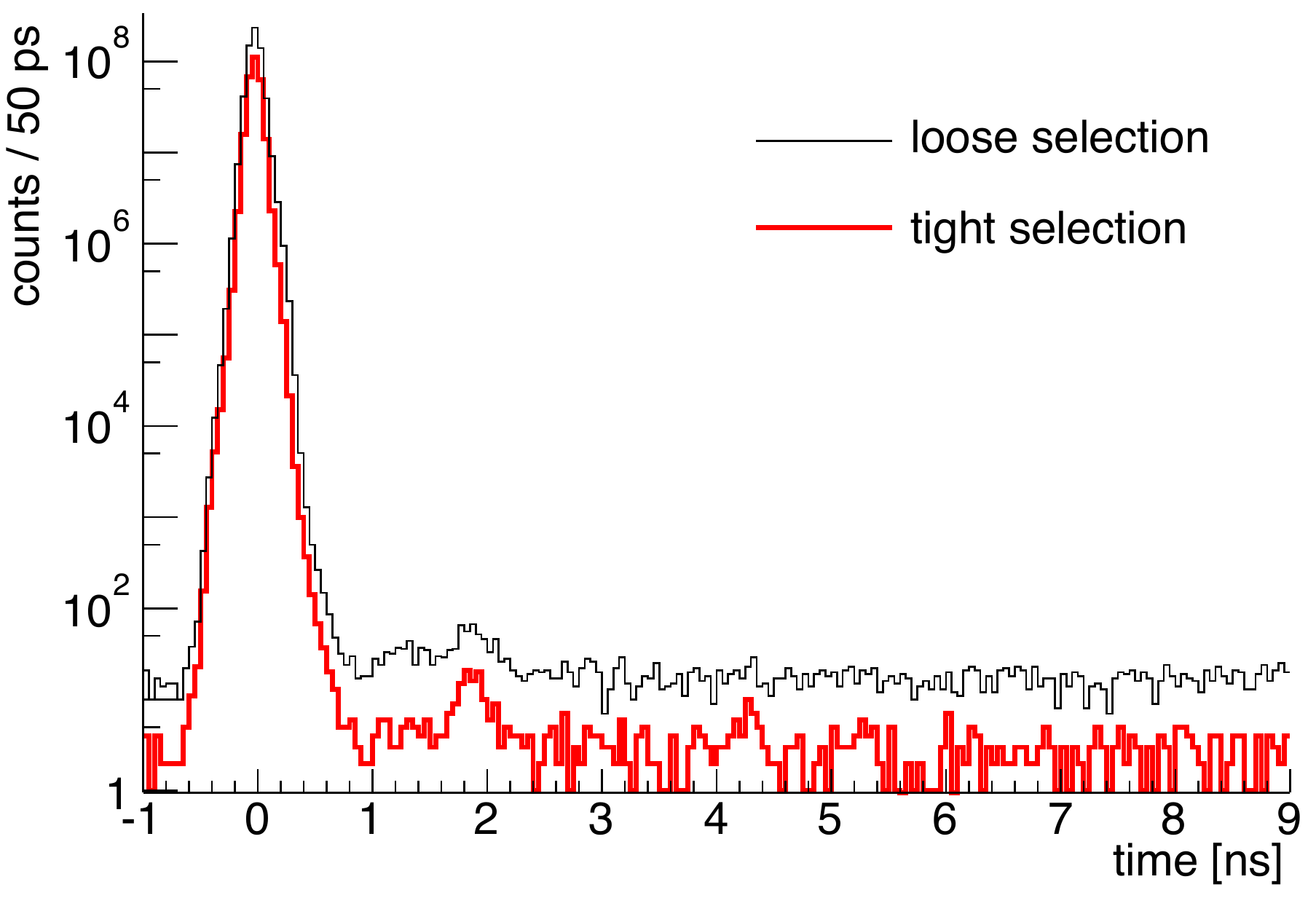}
\caption{\label{fig:thspectrum}(Color online) Time spectra of the search for the NRS from the $^{229}$Th sample. The black solid histogram shows the spectrum with the loose selection and the red bold histogram shows that with the tight selection.}
\end{figure}

 Figure~\ref{fig:thenergy} shows the energy spectra of the prompt scattering and of the constant background. They are obviously different: the spectrum of the prompt scattering consists of the characteristic peaks and that of the radioactive background has no structure around the characteristic peaks. The energy spectrum of NRS signals would also consist of the characteristic peaks of Th, which is the same as the prompt peak, because of its large internal conversion factor of 225;\cite{BROWNE20082657} our system therefore can maximize the experimental sensitivity after the data acquisition by selecting events in an optimum energy range, 12--18\,keV in this case.
 Figure~\ref{fig:thspectrum} shows the time spectra with a loose and tight energy selection; the selected regions are above \SI{5}{keV} (the same as the Hg sample) and 12--\SI{18}{keV}, respectively. The constant background is certainly suppressed and the signal-to-noise ratio after the prompt peak (corresponding to prompt-scattering--to--radioactive-background ratio) increases by a factor of three. We thus successfully demonstrate the system applicability to radioactive nuclides as well as short lifetime measurement. 
 
 We searched for NRS signals by scanning the incident X-ray energy in the range of \SI{\pm11}{eV} around \SI{29.187}{keV} but the obtained spectra did not show any evidence of signals, which is expected to be one order of magnitude smaller than the present sensitivity of the experiment.
 We plan to improve the system, e.g., higher area density of $^{229}$Th sample and higher brightness of the X-ray beam to observe signals.
 
 The \SI{E6}{cps} capability of the energy measurement allows us to accumulate even the prompt scattering data. Because the NRS signal of $^{229}$Th is expected to be located just after the prompt peak within 1--2\,ns due to the short lifetime, it is difficult to ignore the prompt scattering by using a delayed gate. On the other hand, there are many situations in which the prompt scattering can be ignored by a delayed gate.
 A delayed gate ignoring the prompt scattering can allow the system to accumulate only NRS signals up to \SI{E6}{cps} for a channel in such case.
 As the rate capability of Si-APD itself is more than \SI{E8}{cps} in general, this method is feasible in a high rate condition of which the prompt scattering rate is up to \SI{E8}{cps}.
 
\section{Conclusion}\label{sec:conclusion}
 We have developed a fast X-ray detector system for synchrotron radiation based nuclear resonant scattering measurements.
 The system consists of commercially available Si-APDs, fast frontend circuits, and time-to-digital converters.
 The system enables us to acquire both energy and timing information of a single X-ray photon simultaneously with counting rate per channel of more than \SI{1E6}{cps}.
 The performance of the system was investigated in SPring-8. Good time resolution of \SI{120}{ps} (FWHM) and a short tail of $10^{-9}$ level within \SI{1}{ns} were achieved in a high-rate condition of \SI{4.3E6}{cps}. These values are in fastest level among the X-ray detectors used in NRS measurements.
 We have also demonstrated the applicability of simultaneous time--energy measurement by searching for the NRS from the second excited state of $^{229}$Th. By selecting events with an optimum energy range in the post analysis, the signal-to-noise ratio was increased. 
 
The realized performance is important for measurement of very short lifetime resonances. Not only $^{229}$Th, the NRS measurement of the 44.92-keV level of uranium-238 whose half-life is \SI{203}{ps},\cite{toi} for example, is also well motivated.\cite{mossbauer} Uranium-238 is also radioactive nuclide; hence, the energy information is potentially effective for the SR-based NRS measurement of it. This work presents a possibility of expanding the method of a SR-based NRS measurement to a large number of unsurveyed nuclides.\cite{toi}

\begin{acknowledgments}
 The authors thank T.~Taniguchi for designing the CFD, and K.~Inami and K.~Matsuoka for providing the circuit design of the CFD.
 The Th sample was prepared by great cooperation with Y.~Kasamatsu, K.~Konashi, Y.~Shigekawa, M.~Watanabe, and Y.~Yasuda.
 The capillary was provided by T.~Uruga and O.~Sekizawa.
 The authors also thank J.~Tojo and T.~Ito for letting us use a wire bonding machine.
 The synchrotron radiation experiments were performed at the BL09XU beamline of SPring-8 with the approval of the Japan Synchrotron Radiation Research Institute (JASRI) (Proposal Nos. 2014A1334, 2014B1524, 2015B1380, 2016A1420, and 2016B1232).
 T.~H. is supported by Grant-in-Aid for Japan Society for the Promotion of Science (JSPS) Research Fellow.
 This work was supported by JSPS KAKENHI Grant No. 15H03661, Technology Pioneering Projects in RIKEN, and the MATSUO foundation.
\end{acknowledgments}

\bibliographystyle{aipnum4-1}
\bibliography{biblio}

\end{document}